\documentclass[preprint,review,times,3p]{elsarticle}
\usepackage{graphicx}
\usepackage{subfig}
\usepackage[countmax]{subfloat}
\usepackage{amsmath, amssymb}

\biboptions{sort&compress,comma}

\journal{ Physica E}

\begin{document}

\begin{frontmatter}

\title{Temperature Dependent Symmetry to Asymmetry Transition in Wide Quantum Wells}

\author[label1]{G. Oylumluoglu\corref{label4}}
\ead{gorkem@mu.edu.tr} \cortext[label4]{Corresponding Author.
Tel.: +902522111529; Fax: +902522111472} 
\author[label1]{S. Mirioglu}
\author[label2]{S. Aksu}
\author[label1]{U. Erkaslan}
\author[label2]{A. Siddiki}

\address[label1] {Mugla Sitki Kocman University, Faculty of Sciences,
Department of Physics, 48170-Kotekli, \ Mugla, Turkey }
\address[label2] {Department of Physics, Mimar Sinan Fine Arts University, Bomonti 34380, Istanbul, Turkey}

\begin{abstract}
Quasi-two dimensional electron systems exhibit peculiar transport effects depending on their density profiles and temperature. A usual two dimensional electron system is assumed to have a $\delta$ like density distribution along the crystal growth direction. However, once the confining quantum well is sufficiently large, this situation is changed and the density can no longer be assumed as a $\delta$ function. In addition, it is known that the density profile is not a single peaked function, instead can present more than one maxima, depending on the well width. In this work, the electron density distributions in the growth direction considering a variety of wide quantum wells are investigated as function of temperature. We show that, the double peak in the density profile varies from symmetric (similar peak height) to asymmetric while changing the temperature for particular growth parameters. The alternation from symmetric to asymmetric density profiles is known to exhibit intriguing phase transitions and is decisive in defining the properties of the ground state wavefunction in the presence of an external magnetic field, i.e from insulating phases to even denominator fractional quantum Hall states. Here, by solving the temperature and material dependent Schr\"odinger and Poisson equations self-consistently, we found that such a phase transition may elaborated by taking into account direct Coulomb interactions together with temperature.
\end{abstract}

\end{frontmatter}

\section{Introduction}
\label{intro} The interacting quasi-two dimensional (2D) electrons are obtained at the interface of two heterostructures, which have different band gaps. The dimensional constriction yields quantized energy levels and the electron systems are commonly assumed to have zero thickness, i.e. strictly 2D. At low or intermediate doping and at sufficiently low temperatures, only the lowest sub-band is occupied and assuming a $\delta$ function to describe a 2D electron system can be well justified if the resulting quantum well is narrow. In this situation only a single peak is observed at the density distribution in $z$ direction $n_{\rm el}(z)$, which can be approximated by a $\delta(z-z_{\rm el})$. However, the situation becomes quite different if the well is sufficiently wide. Then, the density profile may present more than a single peak, which may have different amplitudes, pointing that also the higher sub-bands are occupied \cite{{Shabani},{Shabani2}}. The effect of surface states and effects due to Coulomb interactions influence the effective potential drastically, together with the fact that the sub-bands become closer in energy \cite{Gerhardts}.
Among many other interesting effects observed at quasi-two dimensional electron systems (2DESs), for instance quantized Hall effects \cite{{Klitzing},{Tsui}}, the observations related with topologically protected ground states attracts attention due to intriguing phase transitions \cite{{Shabani},{Shabani2},{Nayak}}. The states which are claimed to be topologically protected, form at high perpendicular magnetic fields, where the number density of electrons are a fraction of the number density of magnetic flux quantum, so-called the filling factor $\nu$. At even integer dominator filling factors, namely $\nu=3/2,5/2$, quasi-particles are formed due to the many-body interactions. These particles cannot be classified simply as a fermions or bosons due to the uncommon nature of the dimensionality. Hence, braiding statistics has to be utilized which may give Abelian or non-Abelian commutation relations, yielding to topologically protected states \cite{Nayak}. In particular, electron-electron interactions are claimed to be the source of the phase transitions \cite{Shabani2} at wide quantum wells, i.e the phase transition from topologically protected to insulating states.

The interactions are known to be less important in the absence of strong magnetic fields $B$ applied perpendicular to the plane of the 2DES \cite{{Chakraborty},{Oji}}. Once the WQW is subject to a $B$ field, as a rule of thumb to estimate the importance of the interactions one usually compares the distance between these two peaks $d$ in density to the magnetic length $\ell$ ($=\sqrt{\hbar/eB}$). The in-plane correlation energy is inversely proportional to the magnetic length, namely $E_{\mathrm{Corr}}=D e^2/ \epsilon \ell$ where $D$ is a constant at the order of 0.1, and the Coulomb energy is similarly inversely proportional to the peak separation $d$, i.e. $E_{\mathrm{Coul}}\propto e^2/\epsilon d$ \cite{Shayegan}. Hence, the comparison of these two energies together with the symmetric to asymmetric energy gap $\Delta_{SAS}$, determines the properties of the ground state \cite{Shayegan}. It is reported that, the observation of the intriguing fractional states and the formation of insulating phases are strongly affected by the symmetry of these peaks \cite{{Shabani},{Shabani2}}. The experiments show that the even denominator fractional filling factors $\nu=1/2,~1/4$, are present if the density distribution is symmetric and disappears at high imbalance, i.e. density distribution is asymmetric. It is also reported that the insulating phases are observed at low filling factors (e.g. at $\nu=1/5$ \cite{Jiang} considering strong imbalance and, in contrast to even dominator fractional states, are washed out once the system is symmetric \cite{Shabani}. More interestingly, these states are highly temperature dependent. As expected, the fractional states show activated behavior and are characterized by the many-body effects induced energy gap \cite{Shayegan}. The temperature dependency of the activated behavior is strongly influenced by the nature of the wavefunction, i.e. whether the wavefunction is one-component (symmetric density distribution) or two-component (asymmetric density distribution). Another mechanism to change the electron temperature is to derive an external current that increases the electron temperature due to Joule heating. The systematic experimental investigations evidence a melting transition of the insulating phase, where an activated behavior is observed below a certain threshold. This observation is attributed to melting of the Wigner crystal \cite{Shayegan}.

In this letter, we explore the effect of temperature on the density distribution considering a WQW by numerically investigating the band gap variation also taking into account interactions. We utilize the semi-empirical temperature dependent band gap formulation of Varshni \cite{Varshni} and Lautenschlager \cite{Lautenschlager}, and solve the Schr\"odinger and the Poisson equations self-consistently. We show that, depending on the heterostructure parameters one can induce a symmetry to asymmetry transition not only by changing the potential applied to the top or bottom gates, but also by changing the temperature. We propose that, by performing temperature sensitive magneto-transport experiments it is possible to observe a reentrant Wigner crystallization. Such an effect is yet uninvestigated both theoretically and experimentally.

\begin{figure}
{\centering
\includegraphics[width=0.75\linewidth]{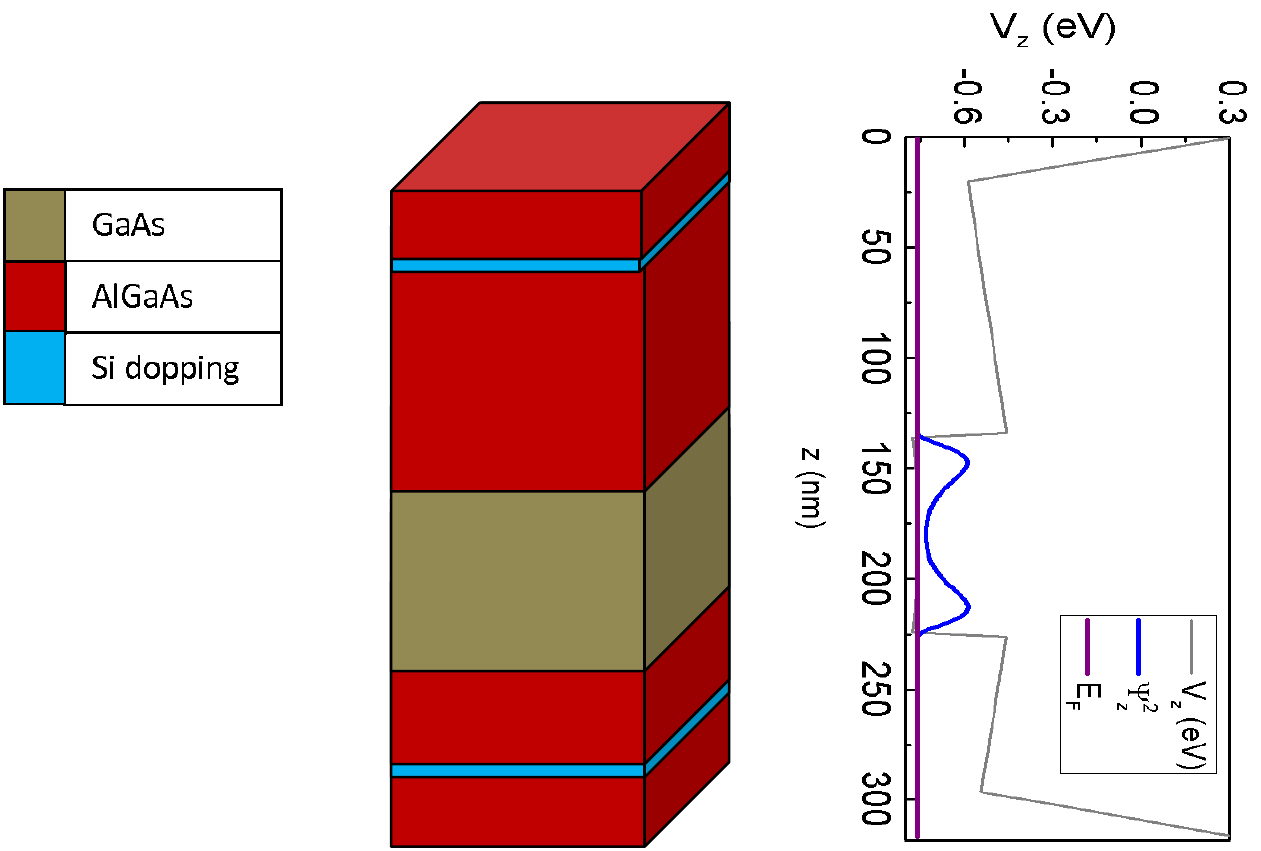}
\caption{\label{fig1}(a) The schematic representation of the heterostructure. (b) The self-consistently calculated conduction band (thin solid line), together with the electron probability distribution $\psi_z^2$ (thick solid line) and Fermi energy (vertical line). The density distribution presents a double peak structure, separated by an average distance $d$.}
}
\end{figure}

\section{The Model}

In fact, solving the Schr\"odinger and Poisson equations in one-dimension considering a quantum well is a straight forward numerical exercise. However, things become complicated if one also takes into account different masses at the well and the barrier, and in addition also the temperature dependency of the energy gap $E_{g}(T)$. In general, such temperature effects emanate from electron-phonon interactions, lattice mismatch (i.e. thermal expansion), etc. The detailed and systematic empirical, numerical and theoretical efforts indicate that, the energy gap is affected by temperature effects even below 1 K \cite{ODonnell}. Despite the fact that there are improved calculation methods \cite{Gonzalez} (and the references therein), the empirical relations proposed by Varshni \cite{Varshni}

\begin{equation} \label{eqn:1}
E_{g}(T)=E(0)-\alpha T^2/(\beta+T),
\end{equation}
and Lautenschlager \cite{{Lautenschlager},{ODonnell},{Gonzalez},{Gopalan}}
\begin{equation} \label{eqn:2}
E_{g}(T)=E(0)-2a_B/(\exp{(\Theta_B/T)}-1),
\end{equation}

are rather simple and also fit the experiments excellently \cite{{Aspnes},{Kim}}. Here, $\alpha$ and $\beta$ are the experimentally obtained constants, whereas $a_B$ is the electron-phonon coupling constant together with the average phonon temperature $\Theta_B$. In our calculations, we utilized both of the relations given above, however, observed that the exact temperature dependency is not decisive. Also recall that the effective mass is almost directly proportional to the energy gap, both for GaAs and Al$_{1-x}$Ga$_x$As heterostructures, as well as the stoichiometry of the heterostructure given by $x$. Keeping these dependencies in mind we solve the Schr\"odinger equation
\begin{equation} \label{eqn:3}
[-\frac{\hbar^2}{2m^\star(z;T)}\frac{d^2}{dz^2}+V(z;T)]\psi(z)=(E-\frac{\hbar^2k^2_{\|}}{2m^\star(z;T)})
\end{equation}
with $E_j(k_{\|})=E-\frac{\hbar^2k^2_{\|}}{2m^\star_j(T)}$ for $j=$w, b, where w and b stands for the well and the barrier, together with the electrostatic potential $V(z,T)$. Hence, the equation yields for $|z|<d/2$
\begin{equation} \label{eqn:4}
\frac{\hbar^2}{m^\star_{\mathrm{w}}(T)}\psi\prime\prime=E_{\mathrm{w}}\psi
\end{equation}
and for $|z|>d/2$
\begin{equation} \label{eqn:5}
\frac{\hbar^2}{m^\star_{\mathrm{b}}(T)}\psi\prime\prime=(E_{\mathrm{b}}-V_0(T))\psi,
\end{equation}
where $V_0$ is the depth of the well, determined by the energy gap difference of the heterostructure. This formulation, allows us to include the effects resulting from both the temperature and the different effective masses. The matching conditions impose that $\psi(z)$ and $\frac{1}{m^\star(z;T)}\frac{d\psi(z)}{dz}$ are continuous, to guarantee the continuity of the electron density $n_{\mathrm{el}}(z)=\int\sum_{n=0}|\psi(z)|^2f(E-E_n,\mu,T)dE$, where $f(\epsilon)$ is the Fermi function, $T$ is the temperature and $\mu$ is the chemical potential. Furthermore, to satisfy the matching conditions, the current density $j_z(z)=\hbar(\psi^{\star}\frac{d\psi(z)}{dz}-\psi\frac{d\psi^{\star}(z)}{dz})/[2im^\star(z;T)]$ across the interfaces and the equation of continuity $\partial n/\partial t+\nabla\cdot \mathbf{j}=0$ should hold. In our calculation scheme we assume that the system is doped by donors, where the donor density is given by $N_D(z)$, and is translational invariant in the $x-y$ plane. Then the total charge density is given by,
\begin{equation} \label{eqn:6}
\rho(z)=-en_{\mathrm{el}}(z)+eN_D(z),
\end{equation}
which generates the electrostatic electric field $E_z(z)=-d\phi(z)/dz$ and the displacement field $D_z(z)=\kappa(z)E_z(z)$, where $\kappa(z)$ is the dielectric constant of the materials. The Poisson's equation can be written as,
\begin{equation} \label{eqn:7}
-\frac{d}{dz}[\kappa(z)\frac{dV_H}{dz}]=-4\pi e^2[n_{el}(z)-N_D(z)],
\end{equation}
where $V_H(z)=-e\phi(z)$ is the Hartree ``potential'' and the total potential energy of an electron is $V(z)=V_0\Theta(|z|-d/2;T)+V_H(z)$. At this point a self-consistent numerical solution is required to obtain the potential and the density given by Eqs.~\ref{eqn:3} and \ref{eqn:7}. For this purpose we employ the numerical algorithm developed by M. Rother, which successfully simulates similar, however, even complicated systems \cite{{Huber},{Huber2}}.

\section{Temperature Dependent Results}

\begin{figure}
{\centering
\includegraphics[width=0.75\linewidth]{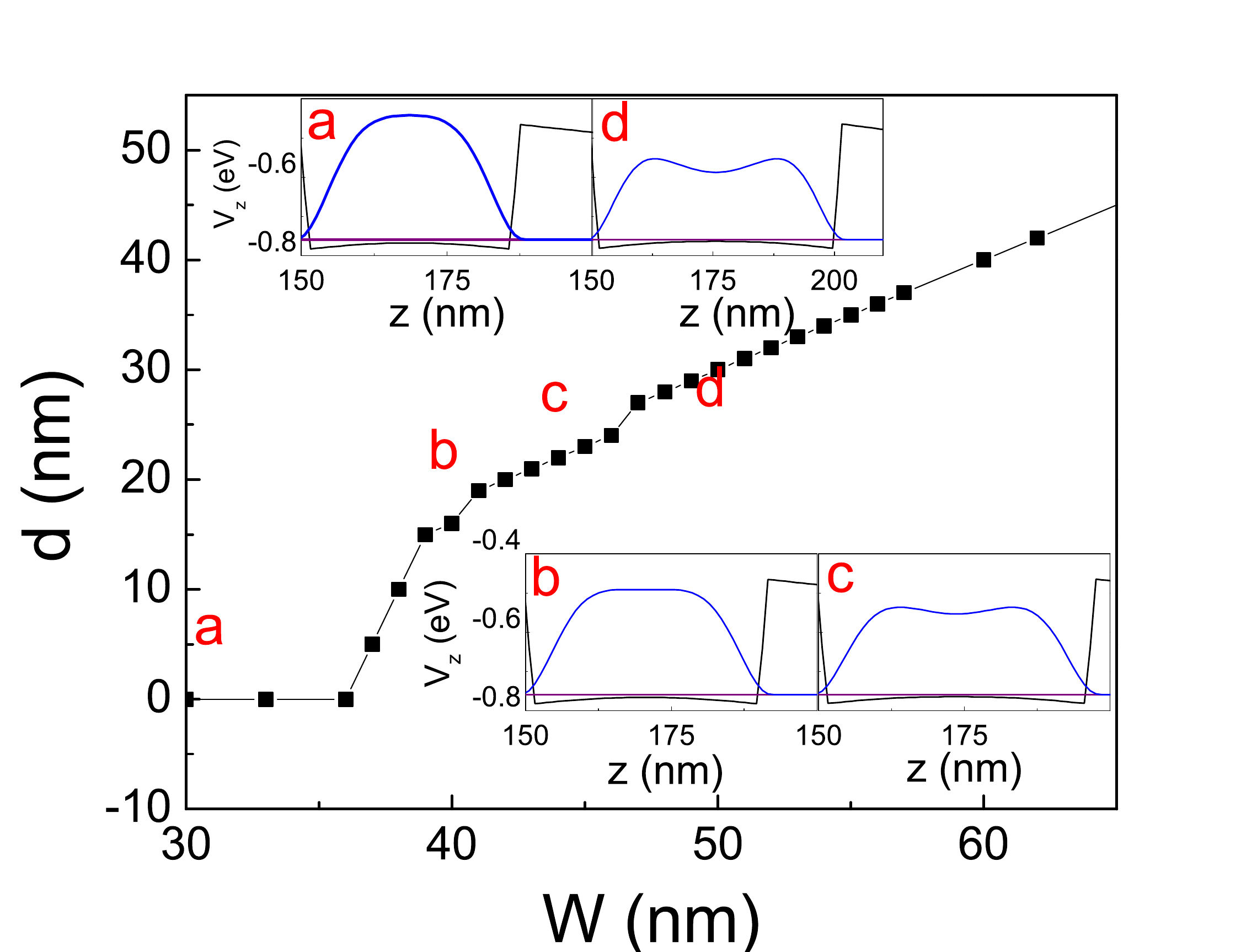}
\caption{\label{fig2}The evolution of peak separation $d$ as a function of well width $W$ at 4.2 K. Insets show density distributions at characteristic $W$. Once the well width is larger than 50 nm, a double peak structure is observed where $d$ scales linearly with $W$.}
}
\end{figure}

Fig.~\ref{fig1}a depicts the schematic presentation of the heterostructure under investigation, whereas Fig.~\ref{fig1}b plots the self-consistently calculated conduction band, together with the probability $\psi^2(z)$ and the Fermi Energy $E_F$ (calculated at $T=0$, otherwise chemical potential $\mu$) as a function of the growth direction. Here the structure parameters are selected such that a double peaked symmetric density distribution is obtained and no top/bottom gates are imposed at the surfaces. To investigate the density distribution as a function of quantum well width $W$, we also performed calculations by varying the thickness of the GaAs material at fixed temperature, namely at 4.2 K, shown in Fig.~\ref{fig2}. We observed that, if the well is narrower than 40 nm only a single peak occurs. Interestingly, when the well width is slightly increased a flat density distribution is obtained within the well. We claim that such a flat, thick electron density distribution yields stable $\nu=1/2,1/4$ states which is still a one-component system. Further increasing the well width essentially results in a linear increase of the peak separation, which presents a symmetric distribution with respect to the center of the quantum well.

\begin{figure}
{\centering
\includegraphics[width=0.75\linewidth]{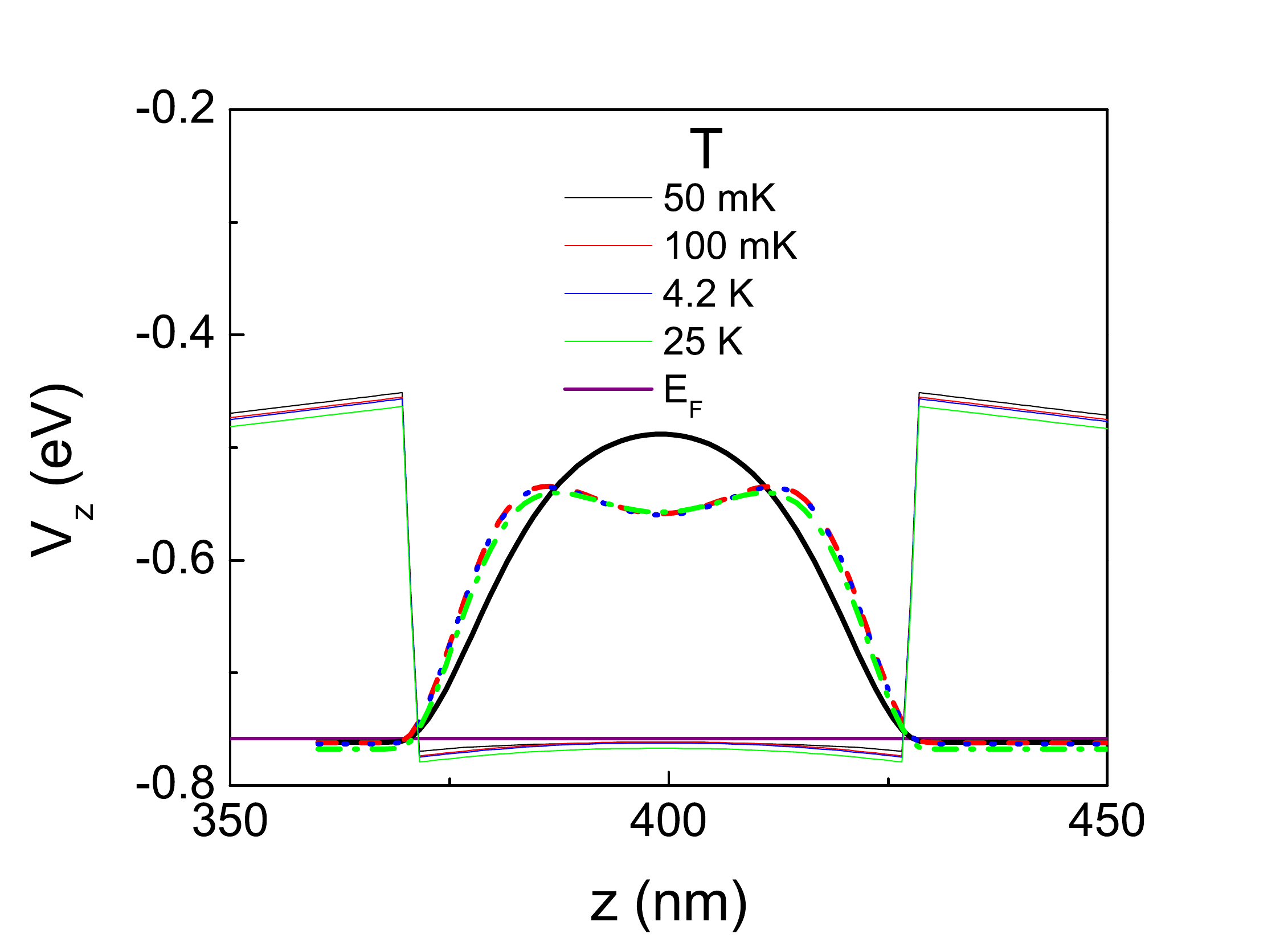}
\caption{\label{fig3}Temperature dependency of an initially symmetric distribution at 50 mK (solid thick line), which evolves to a double peak structure at higher temperatures starting from 100 mK (broken (red) line).}
}
\end{figure}

So far we presented results which are somewhat well known or understood in the existing literature, except the fact that we found a well width interval where the electron density exhibits a constant distribution before two well separated peaks occur. Next, we focus on the effect of temperature on the density distribution. For this purpose, we first start with a symmetrically grown crystal, namely the center of the QW is 400 nm below the surface, where the top (and bottom) 50 nm is capped by a GaAs layer and the 300 nm thick AlGaAs layer is $\delta$doped by Si 70 nm from the surface (and from the bottom) with donor densities of the order of 10$^{19}$ cm$^{-3}$. Fig.~\ref{fig3} depicts the temperature dependency of a symmetric distribution considering a 57 nm wide QW. At the lowest temperature only the lowest sub-band is occupied and we observe a single peak centered around $z=400$ nm. Increasing the temperature from 50 mK to 100 mK results in the occupation of the second level and the double peak structure is observed. Further increase, essentially has approximately no influence on the density distribution, however, the number of electrons within the well is increased, as expected.

\begin{figure}
{\centering
\includegraphics[width=0.75\linewidth]{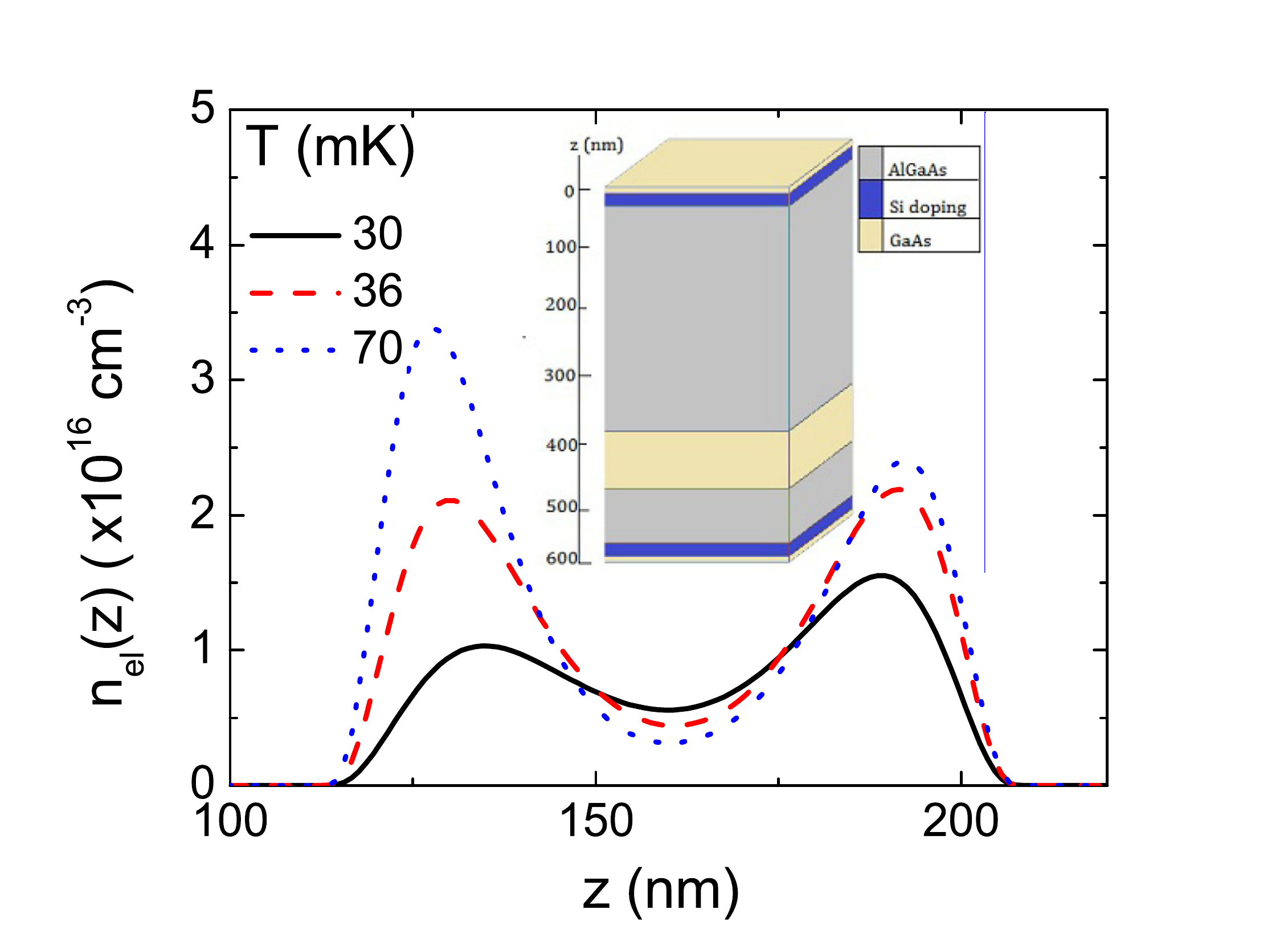}
\caption{\label{fig4}The A-S-A transition while varying the temperature. At the lowest temperature (solid line) ground state is fully occupied, whereas the second level is partially occupied. The electrons are mostly attracted by the lower donor layer, hence, the double peak presents an asymmetry. The distribution is alternated to symmetric at a slightly elevated temperature (36 mK, broken line), and asymmetry is re-established at higher temperatures (70 mK, dotted line).}
}
\end{figure}

This behavior is completely alternated when one already starts with an asymmetric density distribution at lower temperatures. The density asymmetry is obtained by doping the system asymmetrically together with manipulating the distances of the donor layers from the 2DES, as shown in the inset of Fig.~\ref{fig4}. Our main aim is to generate a density imbalance due to different interaction strengths of the electrons and donors, which is not the case for typical experiments. For sure, similar density imbalances can also be obtained by gating the sample, however, we confine our consideration to a situation where charges are fixed by the growth parameters. By doing so, we can eliminate additional effects that may arise due to evaporation. Fig.~\ref{fig4}, shows the evolution of the electron density, while changing the lattice temperature.
At the lowest temperature (solid line), the right side of the WQW is predominately occupied. Note that, in this situation the lowest two sub-bands are already filled with the electrons, however, the next sub-band is merely occupied. The asymmetric locations of the donor layers together with the unequal doping strengths result in different interaction strengths, hence, the electrons are mostly attracted close to the highly doped (lower) donor layer. Once the temperature is increased, the second level is more occupied, however, the extra electrons are repelled to the upper edge of the quantum well, yielding a symmetric density distribution at 36 mK. At the highest temperature the electrostatic equilibrium is established only if the additional electrons are located in the close proximity of the upper side and the density asymmetry is re-constructed.

The observation of asymmetric-symmetric-asymmetric (A-S-A) transition has important consequences on magneto-transport experiments. As mentioned above, if the system remains in a symmetric (balanced) situation, the even integer denominator fractional states are mainly stable. However, we have seen that in an unequally doped system such a stability is limited, hence, observation of fractional states is possible only in a narrow temperature interval, which is still accessible by experimental means. In contrast, the proposed insulating phase can be probed in a large temperature interval, provided that a minimum occurs in the visibility approximately at 36 mK. Such an effect, up to our knowledge, is yet uninvestigated experimentally and we propose that by utilizing unequally doped heterostructures together with varying the well width it is possible to detect this symmetry transition.

\begin{figure}
{\centering
\includegraphics[width=0.75\linewidth]{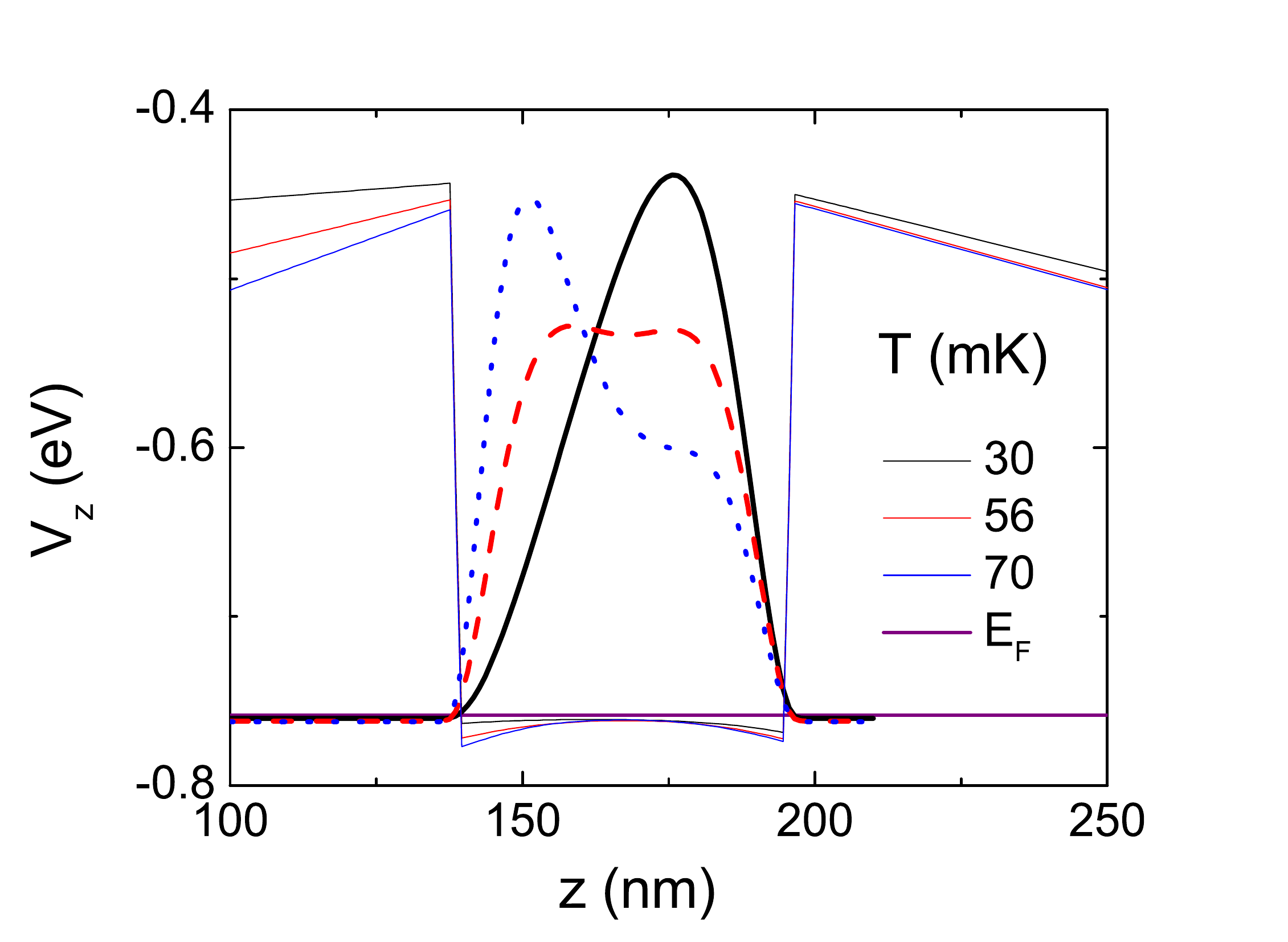}
\caption{\label{fig5}Temperature dependency of the electron density distribution at a relatively narrow quantum well. At 30 mK (solid line) a single peak is observed, which evolves to a symmetric double peak structure at 50 mK (broken line) and to an asymmetric distribution at the highest temperature 70 mK (dotted line).}
}
\end{figure}

\begin{figure}
{\centering
\includegraphics[width=0.75\linewidth]{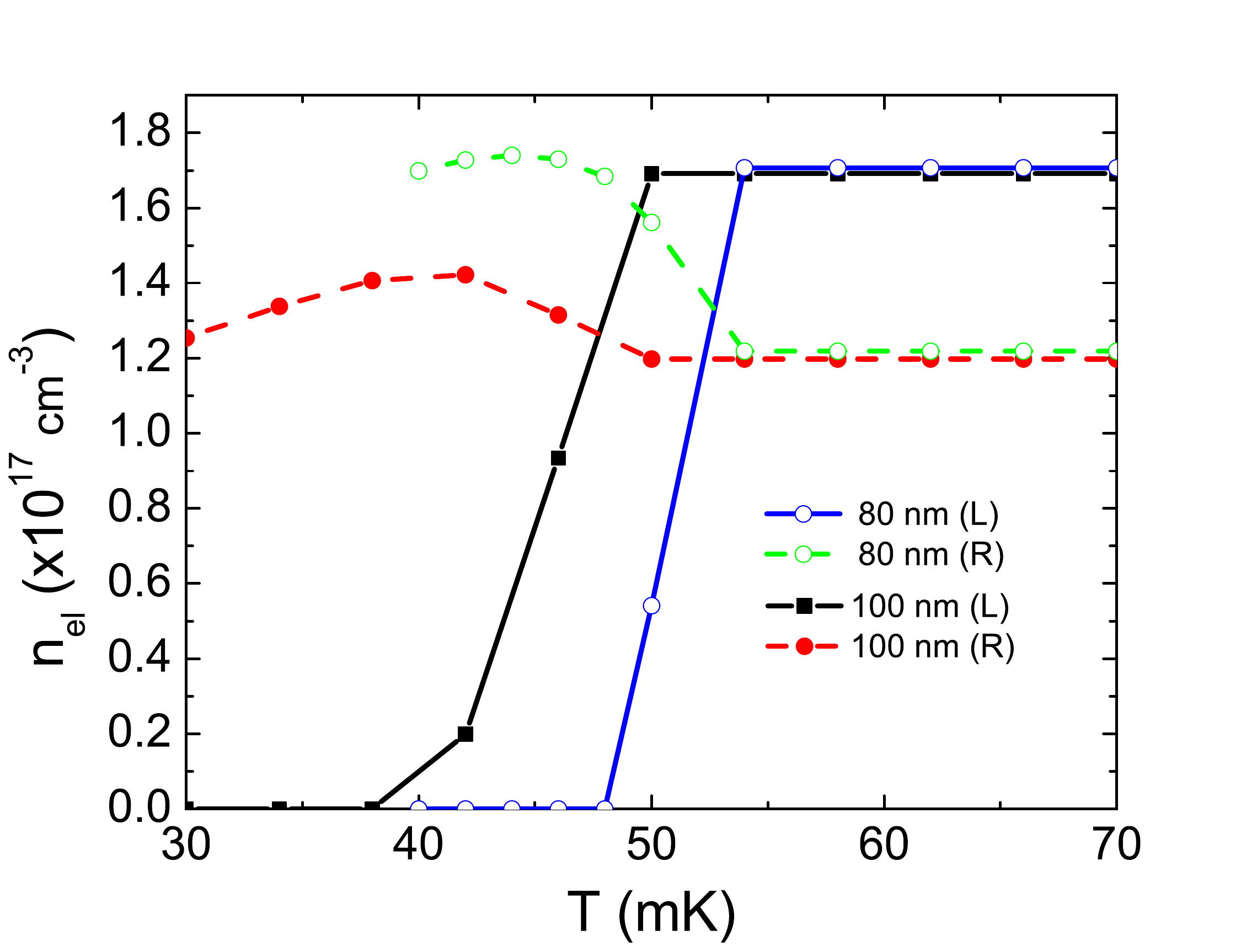}
\caption{\label{fig6}Temperature dependency of the electron density distribution at wide quantum wells, open symbols depict the 80 nm and filled symbols depict a 100 nm wide well. Below 40 mK a single peak is observed for the 100 nm wide well, whereas this temperature is elevated to 50 mK for the 80 nm wide well. the single peak evolves to an asymmetric double peak above 50 mK for both structures.}
}
\end{figure}

In a further step, we change the well width and investigate this transition considering a 57 nm wide well. Our motivation is mainly to simulate the sample structure used by the Shayegan group \cite{Shabani}. Fig.~\ref{fig5} depicts the temperature dependency of the electron density distribution. Similar to the previous case, we observe that the A-S-A transition is still present, however, the system mainly presents a single peak structure, which suppresses the insulating phase transition and enhances the stability of even denominator fractional states. This numerical observation agrees well with the experimental findings that once the electron layer becomes thinner the system presents the properties of a single layer. Hence, our prediction of A-S-A transition can be merely observed at the mentioned experiments. In the opposite limit of a thicker electron layer, the temperature dependent density profile presents the A-S-A transition. This is shown in Fig.~\ref{fig6}, where the peak electron density at left (L) and right (R) are plotted as a function of temperature, for two different widths of the WQW (80 nm, open symbols and 100 nm, filled symbols). One can clearly observe that, there is a critical temperature $T_C$ where the electron densities at different peaks become approximately equal, namely for a 80 nm wide WQW $T_C\simeq 47$ mK and for 100 nm $T_C\simeq 52$ mK. The density mismatch and the temperature intervals compare well with the experiments considering the $\nu=1/2$ \cite{Shabani2}, however, since the well widths and the crystal structures are not compatible, we cannot directly test our results against experiments. To challenge our predictions, further samples should be grown in a controlled and systematic way, in addition precise temperature dependent magneto-transport measurements should be performed.
.

\section{Conclusions}
\label{conc}In this communication we reported on our findings obtained by solving the Schr\"odinger and Poisson equations self-consistently also taking into account the influence of finite temperature on the electron density distribution. We included the effect of temperature both to the occupation function and in addition to the band gap calculations. In particular, we investigated the symmetric to asymmetric transition of the double peaked density distribution considering different growth parameters. It is shown that, if one already starts with a symmetric density distribution within a wide quantum well at low temperatures, the behavior remains unaffected also at elevated temperatures. In contrast, by breaking the symmetry of the growth parameters and starting with an asymmetric density profile at low temperatures, it is observed that the double peak structure goes through a transition, where at intermediate temperatures the profile becomes symmetric. The calculated temperature dependence imposes important consequences on the transport measurements if the 2DES is subject to high perpendicular magnetic fields, such that the symmetric density results in more stable even integer denominator fractional states and may yield a topologically protected ground state, whereas the asymmetric profile imposes that the insulating phase dominates the measurements. Our calculation scheme can be further improved by including the exchange and correlation effects, however, we think that such an improvement would yield only better quantitative results but the qualitative dependency will not be altered.

\bibliographystyle{elsarticle-num}

\end{document}